\documentclass{article}
\usepackage{spconf,amsmath,graphicx}

\title{Improving Emotional Speech Synthesis By Using SUS-Constrained VAE and Text Encoder Aggregation}
\name{Fengyu Yang, Jian Luan, Yujun Wang}
\address{Xiaomi Corporation, Beijing, China}
\begin{document}
%
\maketitle
\begin{abstract}
Learning emotion embedding from reference audio is a straightforward approach for multi-emotion speech synthesis in encoder-decoder systems. But how to get better emotion embedding and how to inject it into TTS acoustic model more effectively are still under investigation. In this paper, we propose an innovative constraint to help VAE extract emotion embedding with better cluster cohesion. Besides, the obtained emotion embedding is used as query to aggregate latent representations of all encoder layers via attention. Moreover, the queries from encoder layers themselves are also helpful. Experiments prove the proposed methods can enhance the encoding of comprehensive syntactic and semantic information and produce more expressive emotional speech.
\end{abstract}
\section{Introduction}
\label{sec:intro}

Emotional speech synthesis is widely applied in various scenarios, such as voice assistant, audio customer service, audio book and etc. Generally, a group of emotion types are defined based on product requirements. For each emotion type, hundreds of sentences are designed and recording lines are collected accordingly. Due to the limited data of each type, usually a multi-emotion TTS model training is implemented by leveraging all the data. To distinguish the data from different emotion types, two questions emerge:1) how to get the emotion embedding from category label; 2) how to inject the emotion embedding into TTS acoustic model. Based on the popular encoder-decoder TTS frameworks, many attempts have been reported to address these two questions.

For the emotion embedding, the basic idea is to learn an embedding vector for each emotion type \cite{LSTM-onehot,Taco-onehot-RNNs}, which is usually called as one-hot embedding or look-up table. However, emotion expression always has slight vibration in intensity or status among utterances, even if the voice talent is guided to keep consistent. Global Style Token(GST) \cite{GST} is then introduced to get utterance-level embedding. At the same time, an emotion classifier can be added to restrict embedding vectors to be clustered by category \cite{GST-classifier}. During inference, the average embedding vector of target emotion type may be employed \cite{GST-avg}, while \cite{GST-interpolation} even tries to control the emotion intensity and inter-emotion transition by interpolating the embedding. Also, GST can generate fine-grained embedding at phoneme level \cite{GST-finegrained}. In recent studies, Variational AutoEncoder(VAE) \cite{VAE} shows stronger capabilities in disentanglement, scaling and interpolation for expression modeling \cite{VAE-expressive} and style control\cite{VAE-style}. However, VAE training is not robust and usually suffers from posterior collapse. 

For the injection of emotion embedding, mostly popular method is to concatenate the embedding vector into decoder input \cite{LSTM-onehot,GST,GST-avg,GST-interpolation,GST-classifier,GST-finegrained,VAE-expressive,VAE-style}, while some studies inject emotion embedding to both attention and decoder RNN layers of Tacotron framework \cite{Taco-onehot-RNNs}. Both of them influence the decoder merely, not considering the effects of emotion embedding to the textual emotion.

In this paper, we propose a framework with innovative solutions for both of the above questions. Firstly, VAE is applied for the emotion embedding. Instead of conventional KL-divergence regularizer, the new constraint expects the means of the embedding vectors are on the surface of the unit sphere while all dimensions have a uniform standard deviation. Secondly for the injection of emotion embedding, this paper uses it as one resource of queries in the attention-based text encoder aggregation to enable emotion-specific sentential information encoding. Another resource of queries are from text encoder themselves as we did in previous study \cite{previous}. In summary, the multi-query attention is designed to capture the syntactic and semantic information better for emotional speech generation. 

\section{Related Work}
\label{sec:rela}

Some previous works also employ VAE embedding as query to attend encoder output. In BVAE-TTS \cite{BVAE}, reference audio is encoded by VAE to get frame-level latent variables. These variables work as query to attend text encoder output for better decoder input. In VARA-TTS \cite{VARA-tts}, however, the output of VAE intermediate layers (called hierarchical latent variables) are all used as query. Different from them, the proposed method encodes reference audio into one vector, i.e. a utterance level emotion embedding, rather than a frame level sequence. Moreover, both BVAE-TTS and VARA-TTS only use the output from the last layer of text encoder as attention memory and the attention is implemented along the time axis. In our framework, the outputs from all intermediate encoder layers are leveraged and the attention is implemented along the stacked layers. In our previous work\cite{previous}, we utilize the contexts extracted form the stacked layers to do self-learned multi-query attention over an expressive corpus. In this paper, we propose to introduce acoustic emotion information to the multi-query for better emotion modeling over a multi-emotion corpus.

\section{Proposed Model}
\label{sec:model}

\begin{figure}[htb]
\vspace{-0.3cm}
  \centering
  \centerline{\includegraphics[width=8.5cm]{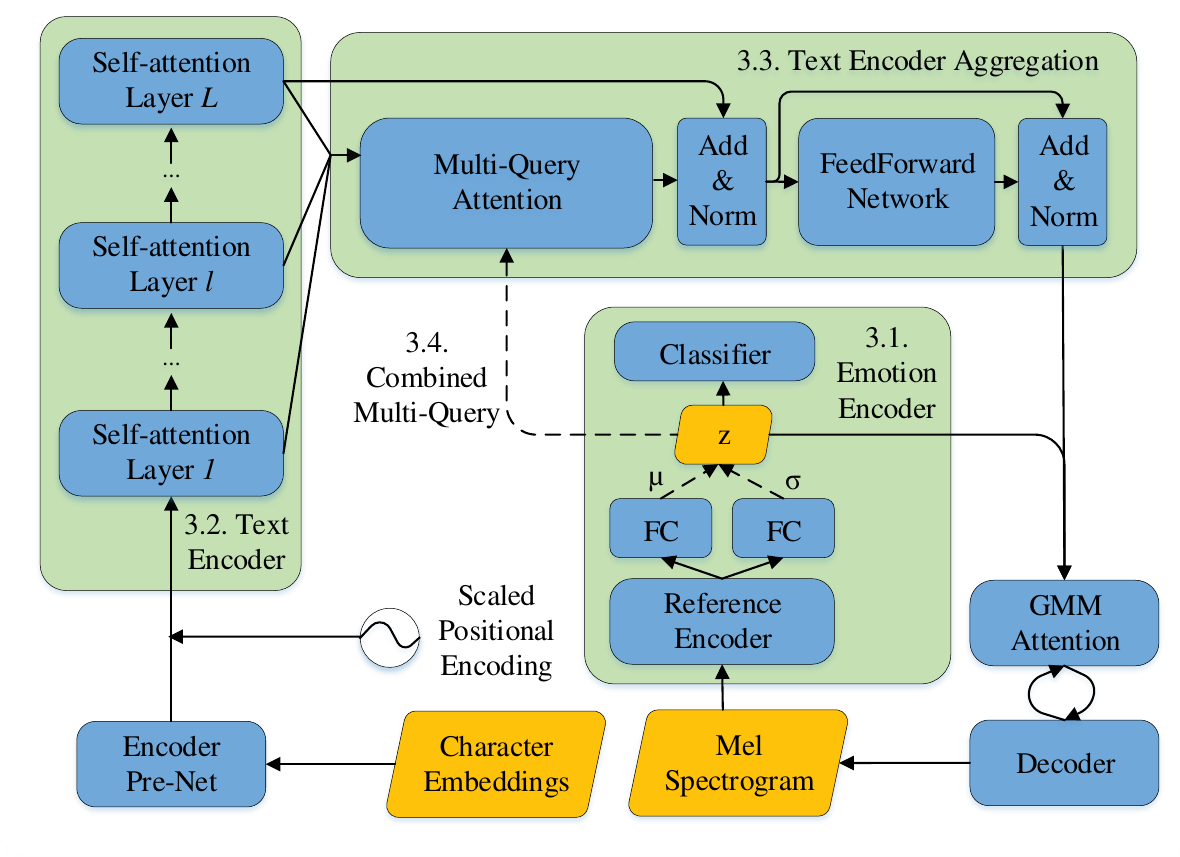}}
  \vspace{-0.4cm}
\caption{Proposed architecture with multi-query attention.}
\label{fig:model}
\vspace{-0.1cm}
\end{figure}

Figure\ref{fig:model} illustrates our proposed approach with text encoder aggregation on exploiting emotional contexts for emotional speech synthesis. It contains a self-attention-based text encoder, an RNN-based auto-regressive decoder, a GMM-based attention\cite{LR-attention} bridging them, a VAE-based emotion encoder and an emotion classifier. WaveRNN\cite{waveRNN} is adopted to convert mel spectrogram to waveforms. The augmented encoder with a context aggregation module will be described in detail.

\subsection{SUS-constrained VAE}
\label{ssec:sus}
VAE does not generate the latent vector directly. Instead, it generates Gaussian distributions each represented by a mean and a standard deviation. During inference, a latent vector is sampled from these distributions. If no additional constraints are employed, the standard deviation will trend to be 0 and sampled latent vectors will always be the mean. Therefore, the desired sampling mechanism becomes invalid. However if a Kullback-Leibler (KL) divergence regularizer is added, it is usually found the generated distributions become normal Gaussian distribution independently of the inputs. Both these two phenomena can be regarded as posterior collapse. Many attempts have been made to address this puzzle, such as annealing strategy in \cite{VAE-style}. 

The critical problem here, we think, is the distances between the means should be in the similar order with their standard deviations. If the distance between means is much bigger than their standard deviation, latent vectors will collapse to the means. Conversely, if the distance between means is much smaller, latent vectors will collapse to be independent on the input. Inspired by it, this paper restricts the means approaching to the Surface of the Unit Sphere (SUS) while set the standard deviations to be an appropriate constant for all dimensions, such as 1. In this way, the distributions of latent vectors will finally have appropriate overlapping proportions, which guarantees VAE's advantages of disentanglement, scaling and interpolation. 

Formally, sampling $z$ from distribution $N(\mu, \sigma^2I)$ is decomposed to first sampling $\epsilon\in(0, I)$ and then computing $z=\mu+\sigma\cdot\epsilon$, where $\cdot$ represents an element-wise product. We restrict the means $\mu$ approaching to the surface of the unit sphere through L2 distance:
\vspace{-0.2cm}
\begin{equation}
  loss_{SUS}=(\sqrt{\sum{(\mu^2)}}-1)^2.
  \label{eq:sus}
\end{equation}
Meanwhile, we set the standard deviations $\sigma$ as a constant.

\subsection{Self-attention based Encoder}
\label{ssec:sa}

Self-attention based sequence-to-sequence framework has been successfully applied to speech synthesis, such as Transformer-TTS\cite{transformer} and Fastspeech\cite{FastSpeech}. We also adopt self-attention networks(SAN) as our based text encoder following \cite{transformer}. Formally, from the previous self-attention block output $H^{l-1}$, the multi-head attention $C^l$ and the followed feed forward network $H^l$ can be computed by:
\vspace{-0.1cm}
\begin{equation}
  C^l={\rm LN}({\rm MH}(head_1^l,\dots,head_H^l)+H^{l-1}),
  \label{eq:C}
\end{equation}
\begin{equation}
  H^l={\rm LN}({\rm FFN}(C^l)+C^l),
  \label{eq:H}
\end{equation}
where MH($\cdot$), FFN($\cdot$) and LN($\cdot$) represent multi-head attention, feed forward network and layer normalization respectively. And in multi-head attention, each head split from the previous self-attention block is calculated as:
\vspace{-0.1cm}
\begin{equation}
  head_h={\rm softmax}(\frac{Q_{h}K_{h}^T}{\sqrt{d}}){\cdot}V_h,
  \label{eq:head}
\end{equation}
where $\{Q,K,V\}$ are queries, keys and values, $d$ represents the hidden state's dimension.

\subsection{Weighted Aggregation}
\label{ssec:wa}

As different SAN layers extract different levels of prosodic-related sentential context information\cite{diff}, we propose a text encoder aggregation module, aggregating them to learn a comprehensive sentence representation to enhance the emotion of the final generated speech. In detail, we utilize a multi-query attention to learn the contribution of each block across the stacked layers. Formally, given a sequence $X$ of $N$ elements, the multi-query calculates the correlation of individual sentential contexts $\{H^0,\dots,H^L\}$ in phoneme level, which are transposed as $\{head_1^g,\dots,head_N^g\}$ for keys and values. We modify Eq.~(\ref{eq:C}) to obtain the weighted contexts:
\begin{equation}
  C^g={\rm LN}({\rm MH}(head_1^g,\dots,head_N^g)+H^L),
  \label{eq:weighted}
\end{equation}
\begin{equation}
  H^g={\rm LN}({\rm FFN}(C^g)+C^g).
  \label{eq:deep-ffn}
\end{equation}
\vspace{-0.5cm}

There are several choices for the multi-query. The first is as our previous work does\cite{previous}, utilizing $\{H^0,\dots,H^L\}$ to obtain the textual multi-query $Q^t$. This self-learned weighted aggregation module leverages the textual contexts information to learn the combination relationship across layers.

\subsection{Combined multi-query}
\label{ssec:co}

For the textual multi-query, the sententail contexts are totally extracted from the stacked textual encoder layers, which does not consider the proved important information from emotion embedding. Commonly the emotion embedding is directly concatenated to the encoder output, influencing the decoder merely. Assuming emotion embedding affects the textual emotion significantly, we propose to introduce contexts extracted from emotion embedding to out multi-query on the basis of direct concatenation.

In details, we employ the output of VAE $vae$ to learn a weighted matrix $Q^a$:
\vspace{-0.2cm}
\begin{equation}
  Q^a={\rm DNN}(vae),
  \label{eq:query-a}
\end{equation}
where DNN($\cdot$) represents a nonlinear transformation with tanh. Then, we combine the contexts information from text and emotion embedding with a learned coefficient to investigate the effectiveness of a comprehensive multi-query:
\begin{equation}
  Q^c=Q^t + {\rm Sigmiod}(w)Q^a,
  \label{eq:query-c}
\end{equation}
where Sigmoid($\cdot$) is a activation function.

\section{Experiments}
\label{sec:exp}

\subsection{Basic setups}
\label{ssec:setup}

To investigate the effectiveness of modeling emotion, we carried out experiments on a Mandarin corpora from a male speaker with 7 emotion categories (neutral, happy, sad, angry, shy, concerned and surprised), which contains about 4.4 hours and a total of 4371 utterances. Except for the neutral, each emotion categories has nearly 500 utterance with consistent emotional strength, separated to non-overlapping training and testing sets (with data ratio 9:1) respectively. For linguistic inputs, we use phones, tones, character segments and three levels of prosodic segments: prosodic word (PW), phonological phrase (PPH) and intonation phrase (IPH). And 80-band mel-spectrogram is extracted from 16KHz waveforms as acoustic targets. For objective evaluation, we conduct mel cepstral distortion (MCD) on test set. And we conduct A/B preference test on 30 randomly selected test set samples with 20 native Chinese listeners as subjective evaluation.

\subsection{Model details}
\label{ssec:setup}

The decoder structure in Tacotron2\cite{Tacotron2} is used as our baseline. But the CBHG encoder and GMMv2 attention are adopted instead for superior naturalness and stability\cite{LR-attention}, where the output of VAE\cite{VAE-style} is added to the encoder output. For the encoder using SAN, the input text embeddings with positional information are pushed to a 3-layer CNN firstly. Then each self-attention block includes an 8-head self-attention and a feed forward sub-network. In the text encoder, there are totally 6 self-attention blocks. As for the aggregation module, $H^L$ is double fed for the convenience of implementation. In VAE module \cite{VAE-style}, the reference encoder consists of six 2D-convolution layers and a GRU layer. Further, the plugged emotion classifier in all systems has a fully connected (FC) layer with ReLu activation and a 7-unit output layer. In our proposed SUS-constrained VAE, we set the standard deviation to 1. WaveRNN is used as vocoder totally following \cite{waveRNN}, trained using the neutral set about 16 hours with the same speaker. For comparison, we built the following different systems:

\begin{itemize}
\item  {\bf BASE}: Baseline system following VAE-Tacotron2~\cite{VAE-style} with CBHG as text encoder and slightly modified GMMv2 attention.
\item  {\bf BASE-SUS}: Baseline system with SUS constraint instead of KL divergence constraint for VAE training described in Section~\ref{ssec:sus}.
\item  {\bf SA-WA}: SAN based encoder with the aggregation module using textual multi-query described in Section~\ref{ssec:wa}.
\item  {\bf SA-WAC}: SAN based encoder with the aggregation module using combined multi-query described in Section~\ref{ssec:co}. (The learned coefficient for combination over the multi-emotion corpus is 0.47.)
\end{itemize}

\subsection{Objective Evaluation}
\label{ssec:obj}

\begin{table}[t]
  \caption{MCD scores of different systems for parallel transfer.}
  \vspace{0.2cm}
  \label{tab:mcd}
  \centering
  \begin{tabular}{c|c|c|c|c}
  \hline
    Emotion & BASE & BASE-SUS & SA-WA & SA-WAC \\\hline
    Neutral & 5.5 & 5.15 & 4.44 & \textbf{4.22} \\\hline 
  \end{tabular}
\end{table}

The MCD results of different systems with parallel transfer are showed in Table~\ref{tab:mcd}. It shows that SUS-constrained VAE has lower MCD than KL constrained VAE, with better ability in generating more similar emotional speech for ground-truth. Meanwhile, it demonstrates that SAN based encoder with text encoder aggregation can improve the performance than the RNN based encoder, where combined multi-query is a better way than textual multi-query in text encoder aggregation to extract emotional contexts. With the help of the injection of the emotion embedding to the textual multi-query, the synthesized speech samples turn into more similar ones to the real speech samples.

\subsection{Subjective Evaluation}

\begin{figure}[htb]
\vspace{-0.4cm}
  \centering
  \centerline{\includegraphics[width=8.5cm]{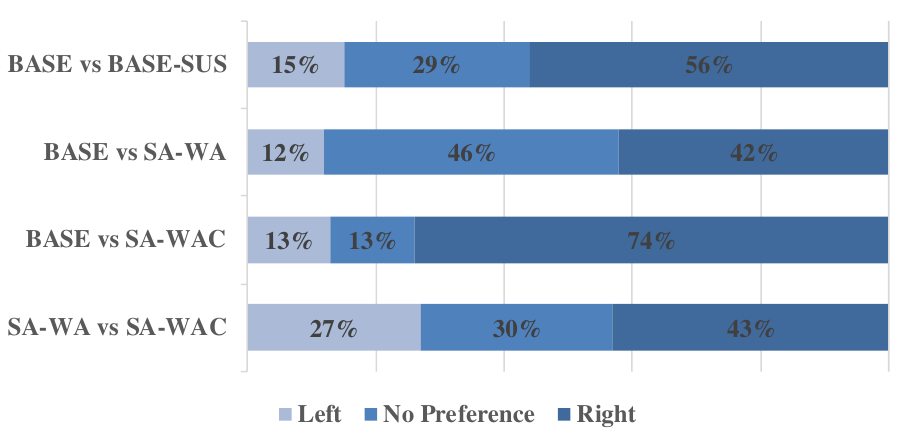}}
  \vspace{-0.4cm}
\caption{A/B preference results for non-parallel transfer with confidence intervals of 95\% and p-value\textless0.0001 from t-test.}
\label{fig:ab}
\vspace{-0.3cm}
\end{figure}

As Figure~\ref{fig:ab} shows, we conduct AB preference tests on the emotional test sets with non-parallel transfer, which include 7 emotion categories. The listeners are asked to select preferred audio according to the overall impression on the expressiveness of emotion in the testing samples\footnote{Samples can be found from: https://fyyang1996.github.io/emotion}. Comparing the BASE system, we find that with the SUS-constrained VAE, our proposed BASE-SUS system obtains much more preferred, due to more expressive emotional speech. Meanwhile, both of the two systems using self-attention based encoder and text encoder aggregation achieve higher preference scores than the RNN based encoder one, which demonstrates that self-attention based aggregation module is a better strategy for generating more expressive emotional speech. Further, combined multi-query attention brings extra performance gain than simple textual multi-query attention according to the A/B preference test.

\vspace{-0.4cm}

\subsection{Analysis}

\begin{figure}[htb]
\vspace{-0.2cm}
\begin{minipage}[b]{.47\linewidth}
  \centering
  \centerline{\includegraphics[width=4.9cm]{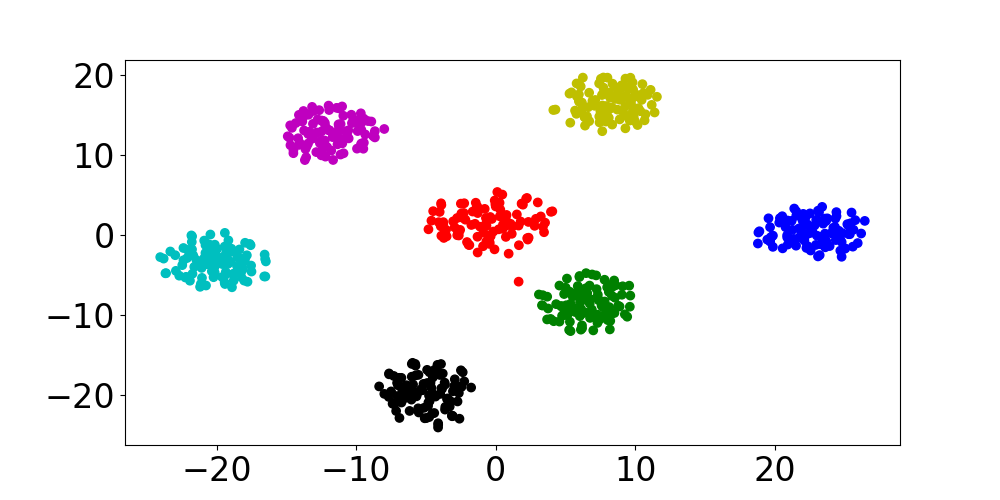}}
  \centerline{(a) The BASE system}\medskip
\end{minipage}
\hfill
\begin{minipage}[b]{.47\linewidth}
  \centering
  \centerline{\includegraphics[width=4.9cm]{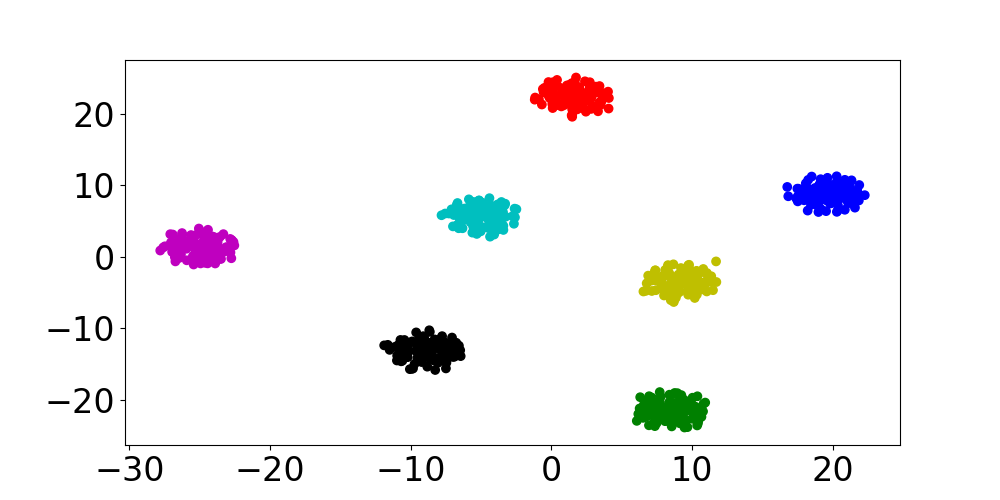}}
  \centerline{(b) The BASE-SUS system}\medskip
\end{minipage}
\vspace{-0.4cm}
\caption{Visualization of two systems using t-SNE for seven emotion categories.}
\label{fig:tsne}
\vspace{-0.3cm}
\end{figure}

\textbf{Emotion Distortion} We visualize the two systems with parallel transfer for seven emotion categories in emotion embedding space by t-distributed stochastic neighbor embedding (t-SNE) plots\cite{tsne}. Figure~\ref{fig:tsne} shows that both the BASE system and the BASE-SUS system appear clear cluster separation, which demonstrates that both two systems have high classification accuracy. But the cluster cohesion of the BASE-SUS system is much better than that of the BASE system. It means that the proposed SUS-constrained VAE can extract emotion information more robustly with less disturbance, which finally helps TTS to generate emotional speech more accurately and expressively. They are also certified in above objective and subjective evaluations.

\textbf{Prosody Correlation} To further estimate the emotion with parallel transfer for statistical significance, phoneme-level intensity, rhythm and intonation of audio are selected. We extract three acoustic features commonly associated with emotion: relative energy within each phoneme (E), duration in ms (Dur.) and fundamental frequency in Hertz (F0) according to \cite{corr, previous}. Additional alignments are done to catch the three prosody attributes in phoneme level. The Pearson correlation coefficient between each system and the ground truth is calculated to evaluate these statistics, using 100 random samples in the test set. The higher Pearson correlation coefficient value demonstrates the higher accuracy of the predicted prosody attribute.

From Table~\ref{tab:correlation} we know that out proposed BASE-SUS achieves higher correlation scores than baseline in all three prosody attributes, which demonstrates that our SUS-constrained VAE has better reconstruction performance in phoneme-level intensity, rhythm and intonation. Meanwhile, in all three prosody attributes, our proposed both SA-WA system and SA-WAC system obtain higher correlation scores than baseline, and SA-WAC system acquires the highest scores. Consequently, we believe that the combined multi-query attention in text encoder aggregation has strong ability in modeling all the three emotional associated attributes.

\begin{table}[t]
  \caption{Correlation in relative energy, duration and F0 within a phoneme computed from different systems for parallel transfer.}
  \vspace{0.2cm}
  \label{tab:correlation}
  \centering
  \begin{tabular}{c|c|c|c|c}
  \hline
     & BASE & BASE-SUS & SA-WA & SA-WAC \\\hline
    E & 0.542 & 0.56 & 0.582 & \textbf{0.595} \\ 
    Dur. & 0.806 & 0.811 & 0.820 & \textbf{0.824} \\ 
    F0 & 0.322 & 0.338 & 0.403 & \textbf{0.422} \\\hline 
  \end{tabular}
\end{table}

\section{Conclusion}

In this paper, SUS-constrained VAE is proposed to extract emotion embedding with better cluster cohesion. Then, based on our previous work of the text encoder aggregation, we introduce the emotion embedding as one resource of queries for attention-based text encoder aggregation. It is with the belief that emotion type should affect the sentential information encoding. Experiments demonstrate that the proposed methods can enhance the encoding of syntactic and semantic information and produce more expressive emotional speech. Moreover, we believe that they can be easily scaled to other tasks, like multi-style or multi-speaker speech synthesis.  


\bibliographystyle{IEEEbib}
\bibliography{strings,refs}

\end{document}